\documentclass[11pt]{article}

\usepackage[margin=1in]{geometry}
\usepackage{amsmath,amssymb}
\usepackage{graphicx}
\usepackage{subcaption}
\usepackage{booktabs}

\usepackage{authblk}
\usepackage{xcolor}
\usepackage[colorlinks=true,allcolors=blue]{hyperref}

\newcommand{\hyperwave}{\textsc{HyperWave}}

\title{\textbf{Learned proposals in trans-dimensional inference are optimal at equilibrium, not during assembly}}

\author[1]{Argyro Sasli}
\author[2]{Nikolaos Karnesis}
\author[3]{Minas Karamanis}
\author[4]{Michael L. Katz}
\author[5]{Dimitrios Kourtesis}
\author[1]{Michael W. Coughlin}
\author[1]{Vuk Mandic}
\author[5]{Nikolaos Stergioulas}
\affil[1]{School of Physics and Astronomy, University of Minnesota, 55455 MN, USA}
\affil[2]{Institute for Astronomy, Astrophysics, Space Applications and Remote Sensing, National Observatory of Athens, 15236 Penteli, Greece}
\affil[3]{Physics Department, University of California and Lawrence Berkeley National Laboratory Berkeley, Berkeley, CA 94720, USA}
\affil[4]{Jet Propulsion Laboratory, California Institute of Technology, Pasadena, CA 91109, USA}
\affil[5]{Department of Physics, Aristotle University of Thessaloniki, Thessaloniki 54124, Greece}

\date{\today}

\begin{document}
\maketitle

\begin{abstract}
\noindent Inferring the dimension of a model --- the number of components needed to explain data --- jointly with the parameters is a pervasive problem, from counting sources in an image to mixture modeling, and reversible-jump Markov chain Monte Carlo solves it exactly but mixes slowly. Learned proposals are well established at fixed dimension, but whether they can accelerate the dimension-changing moves themselves has remained largely untested. We show that the answer has a structural origin: the optimal proposal for the dimension-changing birth move is a different object in different phases of the run. While the fit is being assembled it must match the current residual --- a state-dependent quantity no state-independent network can represent --- but at equilibrium it degenerates to the posterior's single-component marginal, which is exactly the distribution an adaptive normalizing flow learns from the sampler's own history. A learned state-independent birth proposal is therefore useless in one phase and optimal in the other. Controlled experiments confirm the attribution: applied with an exact Metropolis--Hastings correction that leaves the target invariant for any network, the learned births leave acceptance rates unchanged yet accelerate model-order mixing --- in a ten-seed benchmark they meet a pre-specified stopping rule in six of ten runs, typically several times sooner, where a strong hand-tuned baseline meets it in one (one-sided $p=0.03$) --- and an isolation experiment shows the same flow deployed within-model buys nothing. Making no domain-specific assumptions, the same sampler counts sources in a noisy image and reconstructs signals across scientific domains, including gravitational waves from ground- and space-based detectors and a scalp EEG recording. We release the method as \hyperwave{}, an open-source package.
\end{abstract}

\section{Introduction}
\label{sec:intro}

\subsection{Trans-dimensional inference across the sciences}
Inferring the number of components in a model --- and hence the model's dimension --- jointly with the parameters of those components is a recurring problem across the quantitative sciences: the number of change points in a time series, of sources in an image or an interferometric data stream, of sub-populations in a mixture, or of spectral lines in a noisy spectrum~\cite{Green1995,Richardson1997,Green2003}. The Bayesian solution treats the dimension as a random variable and samples the joint posterior over model order and parameters, but doing so efficiently at the data volumes of modern instruments remains a central computational bottleneck. We study this problem through a concrete and demanding instance --- reconstructing an \emph{a priori} unknown transient signal as a variable-size sum of localized time--frequency atoms --- and develop an accelerator that we then show is not tied to that instance.

A leading scientific driver is gravitational-wave (GW) astronomy, where model-free ``burst'' analyses reconstruct a signal without a waveform template. These are indispensable when the source physics is uncertain --- core-collapse supernovae, cosmic-string cusps, post-merger remnants --- or when instrumental glitches must be separated from astrophysical signals~\cite{Cornish2015,Littenberg2015,Klimenko2016}. Two complementary pipelines dominate: coherent WaveBurst (\textsc{cWB})~\cite{Klimenko2016,Drago2021}, a fast coherent excess-power search over a wavelet basis, and \textsc{BayesWave}~\cite{Cornish2015,Cornish2021}, which is fully Bayesian and is the reference point for the present work. \textsc{BayesWave} represents the signal as a variable-size sum of sine-Gaussian (Morlet--Gabor) wavelets and samples the joint posterior over the wavelet count $D$ and their parameters. Its reconstruction fidelity is well established --- a network match of $\sim0.94$ with the numerical-relativity waveform of GW150914~\cite{Abbott2016,Abbott2016b,Abbott2019gwtc1,Ghonge2020} --- and it underpins detection-confidence, sky-localization, glitch subtraction, and tests of general relativity~\cite{Abbott2016tgr} across the LIGO--Virgo--KAGRA~\cite{Aasi2015,Acernese2015} observing runs. The computational cost, however, is substantial, and the coming era of the Laser Interferometer Space Antenna (LISA) and next-generation ground detectors demands both higher throughput and applicability across frequency bands and instruments that differ by orders of magnitude.

\subsection{The statistical machinery and its accelerators}
The natural sampler for a signal of unknown dimension is reversible-jump MCMC (RJMCMC)~\cite{Green1995,Green2003}, which augments within-model moves with dimension-changing birth and death moves under a dimension-matching Jacobian. Trans-dimensional sampling is notoriously slow to mix: birth moves that place a new component blindly are almost always rejected, and the parameters of a faint component are only weakly constrained. Two families of accelerators have been pursued. (i)~\emph{Data-informed proposals} build the birth distribution from the whitened residual (a matched-filter or $\mathcal{F}$-statistic map), as in the LISA global-fit literature~\cite{Littenberg2020,Katz2024}; these target what is currently \emph{missing} from the fit. (ii)~\emph{Learned proposals} fit a flexible density --- historically a Gaussian mixture, now a normalizing flow --- to the accumulating chain and propose from it. Normalizing flows~\cite{Rezende2015,Papamakarios2021} --- in particular masked autoregressive~\cite{Papamakarios2017} and neural spline flows~\cite{Durkan2019} --- provide exact densities and fast sampling, and have been used both as MCMC proposals~\cite{Gabrie2022pnas} and as the transport map inside preconditioned sequential Monte Carlo (\textsc{pocoMC})~\cite{Karamanis2022}.

\paragraph{Relation to existing flow-enhanced samplers in gravitational-wave inference.}
The mechanism we use --- an adaptive normalizing flow, retrained online from the chain and applied as an independence proposal under an exact Metropolis--Hastings correction --- is the same mechanism that underlies \textsc{flowMC}~\cite{Wong2023flowMC} --- which implements the adaptive-flow scheme of Ref.~\cite{Gabrie2022pnas} --- and its GW application \textsc{jim}~\cite{Wong2023jim}, and is closely related in spirit to the flow-augmented nested sampling of \textsc{nessai}~\cite{Williams2021nessai,Williams2023ins}. Our contribution is therefore not the mechanism itself but its extension to, and characterization within, the \emph{trans-dimensional} setting: all of these samplers operate at fixed, known parameter dimension, where the flow models a single target density. When the dimension is itself a random variable, the sampler alternates between within-model moves --- for which the flow is well posed --- and dimension-changing moves, for which, as we show in Sec.~\ref{sec:results-births} and Methods~\ref{sec:births}, the value of a \emph{state-independent} learned density is phase-dependent: useless during assembly, decisive for model-order mixing at equilibrium. Making that division of labor explicit, and quantifying what it buys, is the point of this paper.

A second observation follows: a flow learns the chain's component distribution, which is the wrong object for a birth \emph{while the fit is being assembled} (that phase requires the state-dependent residual) but --- as we show --- exactly the right object once the chain is at equilibrium, where matching the posterior's component marginal is what drives model-order mixing.

\subsection{Amortized and simulation-based inference}
A complementary line replaces per-event sampling with amortized inference: a neural conditional density $q_\varphi(\theta\,|\,d)$ is trained once on simulations and evaluated in milliseconds at test time~\cite{Cranmer2020,Papamakarios2019}. In GW astronomy this is realized by \textsc{Dingo}, which attains posterior estimates for compact-binary coalescences in seconds~\cite{Dax2021,Dax2023}. Amortized inference is, however, almost always \emph{fixed}-dimensional: the parameter vector has a known, constant size. Amortizing a \emph{trans}-dimensional posterior --- predicting both the number of components and their parameters --- has only very recently been attempted: \textsc{SlotFlow}~\cite{Houba2025slotflow} pairs a component-count classifier with slot-based conditional flows on sinusoidal decompositions of up to ten components. Such one-shot amortization trades exactness for speed --- the posterior is only as accurate as the network --- and is complementary to the present work, where sampling remains exact and learning enters only through proposals. Calibration of amortized posteriors is assessed with simulation-based calibration and coverage tests~\cite{Talts2018}.

\subsection{This work}
\begin{figure}[!tbp]
\centering
\includegraphics[width=\linewidth]{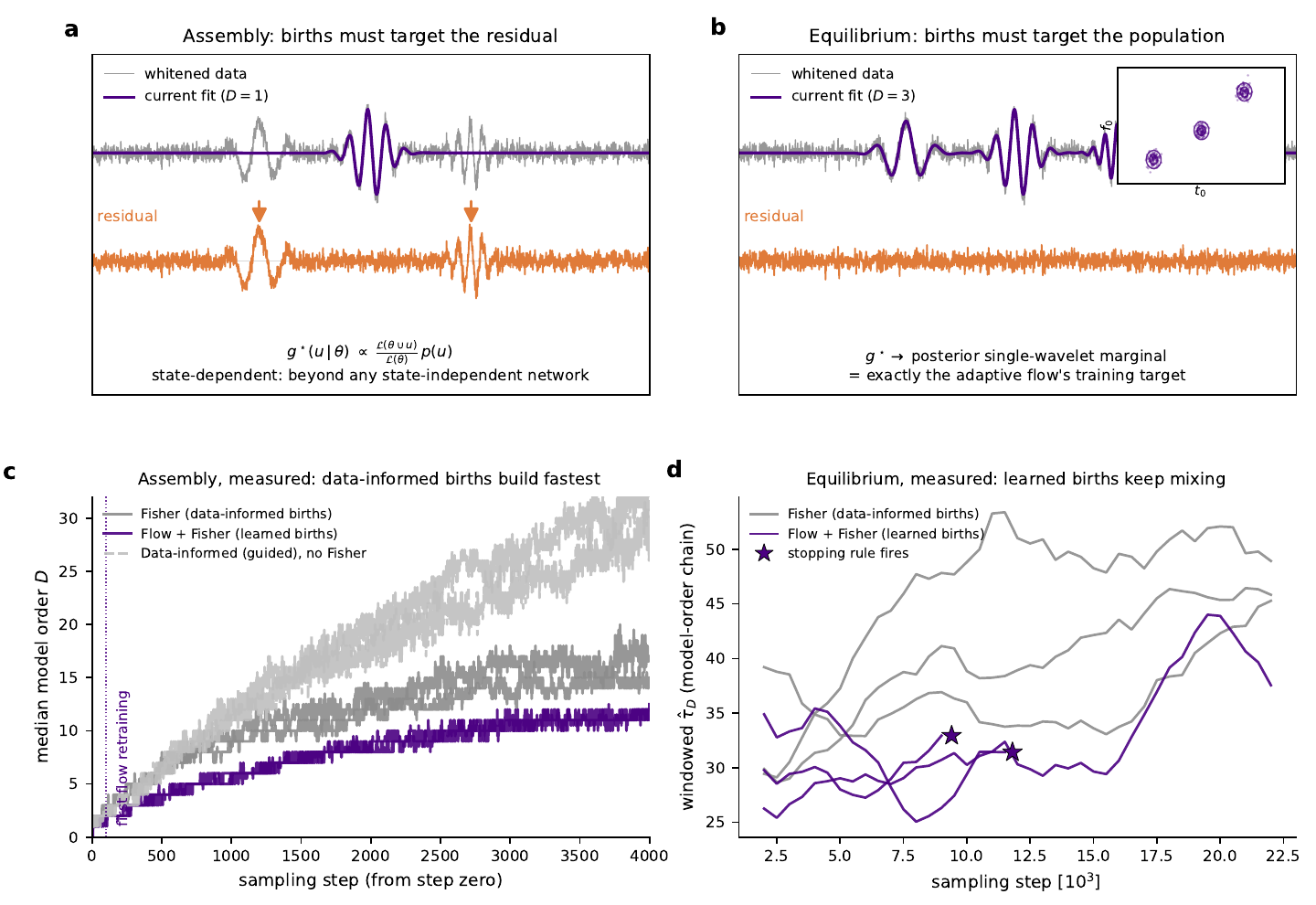}
\caption{\textbf{The phase structure of the birth proposal.}
\textbf{a}, While the fit is being \emph{assembled}, the whitened residual (orange) carries the structure the model has not yet captured, and the optimal birth proposal is the residual-matched, state-dependent distribution $g^\star(u\,|\,\theta)$ of Eq.~\eqref{eq:gstar} (arrows: where births must be placed). No state-independent learned density can represent this object.
\textbf{b}, At \emph{equilibrium} the residual is statistically pure noise and $g^\star$ degenerates to the posterior's single-wavelet marginal (inset: the population of wavelets in the chain, with the adaptive flow's density contours) --- exactly the distribution the flow is trained on, so the learned birth becomes optimal precisely where hand-designed proposals run out of signal to match.
\textbf{c}, The assembly phase, measured: median model order against sampling step from step zero (no burn-in discarded; two seeds per configuration). Data-informed births assemble the fit fastest (grey); learned births (indigo) start from the defensive prior fallback and assemble visibly more slowly --- the predicted assembly-phase deficit of a birth proposal that does not target the residual --- while the guided configuration without in-model refinement (light grey, dashed) compensates poor fits by inflating $D$ toward the prior ceiling, the ``reduced fidelity'' row of Table~\ref{tab:bench}.
\textbf{d}, Equilibrium, measured: the stopping statistic itself --- the two-half total-variation distance of $p(D)$ --- against sampling step, for all twenty runs of the benchmark campaign of Table~\ref{tab:bench} (ten seeds per configuration; a trajectory ends where its run fired the rule, marked by a star, or at the $72$k cap). With learned births (indigo) the statistic reaches the pre-specified threshold (dashed) in $6/10$ runs, at $12$--$49$ thousand steps; with data-informed births (grey) it does so in $1/10$. Panels a and b are schematic; panels c and d are measured.}
\label{fig:phases}
\end{figure}

We make three contributions, released together as the open-source \hyperwave{} package.
\emph{(1)}~The main result is structural (Fig.~\ref{fig:phases}): the optimal proposal for the dimension-changing birth move has a \emph{phase structure}. While the fit is being assembled it is the residual-matched, state-dependent distribution of Eq.~\eqref{eq:gstar}; at equilibrium it degenerates to the posterior's single-component marginal --- exactly the distribution an adaptive flow trained on the chain's own history learns. A state-independent learned birth is therefore useless in one phase and optimal in the other --- which explains at once why such a proposal can fail to help and when it pays off, including the negative results we report ourselves. Controlled and isolation experiments confirm the attribution: learned births leave acceptance rates unchanged yet accelerate model-order mixing, meeting a pre-specified stopping rule in $6/10$ seeds --- typically several times sooner --- where a strong hand-tuned baseline meets it in $1/10$ (one-sided $p=0.03$), while the same flow deployed as an \emph{in-model} independence move buys nothing. \emph{(2)}~The machinery: an adaptive neural spline flow retrained online from the cold chain under an exact Metropolis--Hastings/Green correction --- the target is invariant for \emph{any} network, so no validation of the flow is ever required --- combined with a batched, GPU-resident likelihood kernel that evaluates the entire tempered ensemble in a single device call. \emph{(3)}~Generality: with no domain-specific assumptions, the identical sampler counts 2-D pulses in a noisy image, reconstructs GW150914 from public LIGO data, a simulated LISA massive black-hole binary, and a real electroencephalogram, and applies unchanged to \emph{fixed}-dimensional template-based inference. \hyperwave{} is built as an off-the-shelf tool: any likelihood plus a component template defines a new application (Methods~\ref{sec:interface}). All results are reproducible with the package.

\section{Results}
\label{sec:results}

\subsection{Benchmark: what the learned proposal buys}
\label{sec:results-bench}

Table~\ref{tab:bench} reports the head-to-head benchmark on a common injection --- a compact-binary coalescence (\textsc{IMRPhenomPv2}, component masses $\sim36/29\,M_\odot$, $600$~Mpc) injected into simulated H1/L1 Gaussian noise over $[20,512]$~Hz, network SNR $45.8$, with the sky sampled --- with \emph{ten} random seeds per configuration, run under the pre-specified stopping rule of Methods~\ref{sec:converge} with a hard cap of $72\,000$ sampling steps. The cap follows the horizon analysis of Methods~\ref{sec:converge}: the stationarity statistic carries a sample-size noise floor, and a shorter budget cannot resolve the rule for \emph{any} proposal --- a point we learned the hard way (Methods~\ref{sec:repro}). All configurations are initialized identically --- walker coordinates drawn from the priors under the same per-seed random state --- and differ only in their proposal kernels. Throughout, \emph{overlap} denotes the noise-weighted network match to the injection, Eq.~\eqref{eq:match}, and \emph{in-model acceptance} the acceptance fraction of fixed-$D$ moves on the cold chain (Methods~\ref{sec:notation}).

The result is a clean separation in \emph{when the stopping rule is reached}, at identical reconstruction fidelity. The flow-assisted cascade satisfies the rule in $6$ of $10$ seeds, at $12.0$--$49.2$ thousand steps (median $29.7$k; the four remaining seeds end the cap at $\mathrm{TV}=0.024$--$0.070$); the Fisher-only cascade satisfies it in $1$ of $10$ (at $61.4$k, the rest capped at $\mathrm{TV}=0.022$--$0.040$). A one-sided Fisher exact test on the binary outcome gives $p=0.029$. Figure~\ref{fig:phases}d shows the full time-resolved stationarity statistic for all twenty runs. The mixing difference survives the length-matched comparison, which is the controlled one because automatic-windowing estimators of $\hat\tau$ grow with chain length: among the runs that reach the identical $72$k cap, the flow configuration's $\hat\tau$ spans $51.3$--$55.3$ (four runs) against $61.9$--$78.2$ for the Fisher cascade (nine runs) --- fully disjoint ranges. As in every experiment we report, the reversible-jump \emph{acceptance} is left essentially unchanged by the learned births ($\approx0.09$--$0.11$ for both configurations), while the in-model acceptance of the \emph{identical} Fisher cascade rises indirectly ($0.152\to0.167$ at the median), reflecting better-mixed states. Reconstruction fidelity is the same wherever it can be compared: median network overlap $0.954$ for the flow-assisted runs at their stopping points, and $0.930$--$0.931$ for both configurations at a matched $24$k-step length. The claim is therefore not ``better reconstructions'' but ``a pre-specified stopping rule that is actually reached, several times sooner and in six times as many seeds.''

\begin{table}[t]
\caption{Head-to-head benchmark on a common injection (network SNR $45.8$; \emph{ten} random seeds per configuration; $72\,000$-step cap; NVIDIA V100 GPUs). \emph{Steps to rule} is the sampling-step count at which the pre-specified stopping rule of Methods~\ref{sec:converge} was satisfied (distribution over the seeds that reached it); at the benchmark's sampling cost of $\approx50$~ms per step, $10^4$ steps $\approx8$~min of wall clock. $\hat\tau$ at cap is the integrated autocorrelation time over the runs that reached the identical $72$k length --- the length-matched comparison. \emph{In-model acc.} is the acceptance fraction of fixed-$D$ moves on the cold chain (medians); the reversible-jump acceptance is statistically identical for the two configurations ($\approx0.09$--$0.11$). No diagnostic certifies convergence; the rule is a pre-specified decision procedure applied identically to both arms.}
\label{tab:bench}
\centering
\begin{tabular}{lcccc}
\toprule
Proposal & Rule met & Steps to rule & $\hat\tau$ at cap & In-model acc. \\
\midrule
Fisher                     & $1/10$ & $61.4$k & $61.9$--$78.2$ \,($n{=}9$) & $0.152$ \\
\textbf{Flow\,+\,Fisher}   & $\mathbf{6/10}$ & $\mathbf{12.0}$--$\mathbf{49.2}$\textbf{k (med.\ 29.7k)} & $\mathbf{51.3}$--$\mathbf{55.3}$ \,($n{=}4$) & $\mathbf{0.167}$ \\
\bottomrule
\end{tabular}
\\[3pt]
{\footnotesize One-sided Fisher exact test on the rule-met outcome: $p=0.029$. Fidelity is identical where comparable: flow-assisted median network overlap $0.954$ at the stopping points; $0.930$ against $0.931$ for the two configurations at a matched $24$k length. Non-converged runs ended the cap at $\mathrm{TV}=0.024$--$0.070$ (flow) and $0.022$--$0.040$ (Fisher).}
\end{table}

\subsection{Where the mechanism helps, and where it does not}
\label{sec:results-births}

The benchmark of Sec.~\ref{sec:results-bench} localises the gain in an unexpected place, and the acceptance rates explain why it was easy to mis-attribute. Reusing the trained flow as the birth proposal $g$ leaves the reversible-jump \emph{acceptance} essentially unchanged ($0.09$--$0.11$ across all birth proposals we tested), which suggests --- wrongly --- that learned births are useless. What changes is the \emph{mixing of the model-order chain}: with flow births the $p(D)$ stationarity criterion is met in $6/10$ seeds and typically several times sooner, while with data-informed births it is met in $1/10$, whether or not an in-model flow is added (Table~\ref{tab:bench}, Fig.~\ref{fig:phases}d).

The resolution is that the optimal birth proposal is a different object in different phases of the run. While the fit is being \emph{assembled}, the Bayes-optimal birth is proportional to the likelihood gain of adding one wavelet, $g^\star(u\,|\,\theta)\propto\mathcal{L}(\theta\cup u)/\mathcal{L}(\theta)$, which depends on the current residual $r=d-h(\theta)$: births must target what is missing, and the data-informed proposal approximates exactly this. Once the fit is assembled and the chain is \emph{at equilibrium}, however, birth--death exchange is fastest when the birth proposal matches the posterior's single-wavelet marginal --- and that marginal is precisely the distribution the adaptive flow learns from the chain. A state-independent learned density is therefore the wrong object during build-up but the right object at stationarity; the acceptance rate, which averages over both phases and is dominated by the many rejected proposals either way, resolves none of this. We state plainly what the benchmark does and does not test: no benchmarked configuration switches proposals between phases --- the winning configuration uses learned births throughout, falling back to the prior before the flow's first training --- so the head-to-head rows of Table~\ref{tab:bench} directly probe the equilibrium half of the argument, where the run spends nearly all of its steps. The assembly half is measured separately: run from step zero, data-informed births assemble the fit markedly faster than learned births (median $D\approx4$--$5$ against $2$--$3$ at step $400$; Fig.~\ref{fig:phases}c) --- precisely the assembly-phase deficit the argument predicts for a birth proposal that does not target the residual. The sharpest untested prediction of the two-phase picture is a \emph{switched} proposal --- data-informed births until the residual is noise-dominated, learned births thereafter --- which should outperform both pure configurations; we return to this in the Discussion.

The same accounting explains the in-model side. The in-model acceptance rise in Table~\ref{tab:bench} ($0.152\to0.167$) occurs with \emph{identical} Fisher cascades and is an indirect signature of better-mixed states, not a direct effect of a learned in-model move; deployed directly in-model (the isolation experiment below), the flow independence proposal \emph{lowers} the cascade's acceptance and does not improve convergence, consistent with the fixed-dimensional result of Sec.~\ref{sec:fixed-pe}.

\paragraph{Ablations and variants.} Further experiments complete the attribution. \emph{(i) Isolation:} data-informed births plus an \emph{in-model} flow independence move (weight $0.3$) lowers the in-model acceptance to $\approx0.10$ and leaves model-order stationarity unimproved ($0/13$ across two hardware generations) --- the gain tracks the \emph{birth} column, not the in-model column. \emph{(ii) Q-marginal:} a natural alternative explanation is that the flow simply learns the posterior of the quality factor $Q$, which the data-informed birth draws from its prior over a $400$-fold range. Replacing \emph{only} that draw with a marginal refit online from the cold chain (a floored histogram; density estimation, no flow) and running the identical ten-seed, $72\,000$-step protocol gives $2/10$ rule-met against $1/10$ for the unmodified baseline and $6/10$ for the flow --- statistically indistinguishable from the baseline ($p=0.5$), and below the flow. Learning one nuisance marginal is therefore not the mechanism; what the learned birth contributes is the \emph{joint} five-dimensional structure of the component population, not any single margin. \emph{(iii) Multiple-try births:} proposing four flow candidates per birth and selecting by likelihood doubles the reversible-jump acceptance ($\approx0.09\to0.18$--$0.20$) and removes the flow's assembly deficit (the fit starts at $D\approx17$), at four likelihood evaluations per birth --- a useful lever where the likelihood is cheap. \emph{(iv) SNR-weighted deaths} (deaths preferring low-amplitude components, exactly corrected) churn the low-amplitude population without improving stationarity --- an instructive negative. Ablations (iii) and (iv) were run at a $24\,000$-step cap, at which --- per the noise-floor analysis of Methods~\ref{sec:converge} --- no configuration resolves the stopping rule; they are therefore compared on acceptance and $\hat\tau$ only. Finally, in a \emph{crowded} variant of the problem ($K=30$ overlapping wavelets), assembly dominates any budget we could afford and the data-informed births lead throughout --- the assembly half of the phase argument, again. A flow conditioned on a summary of the current residual would interpolate between the two birth regimes by construction and remains, in our view, the natural next step (Sec.~\ref{sec:conclusions}); general frameworks for valid dimension-changing kernels~\cite{Neklyudov2020,Green2003} accommodate such maps.

\subsection{The posterior recovers the true model order}
\label{sec:modelsel}

\begin{figure}[!tbp]
\centering
\includegraphics[width=\linewidth]{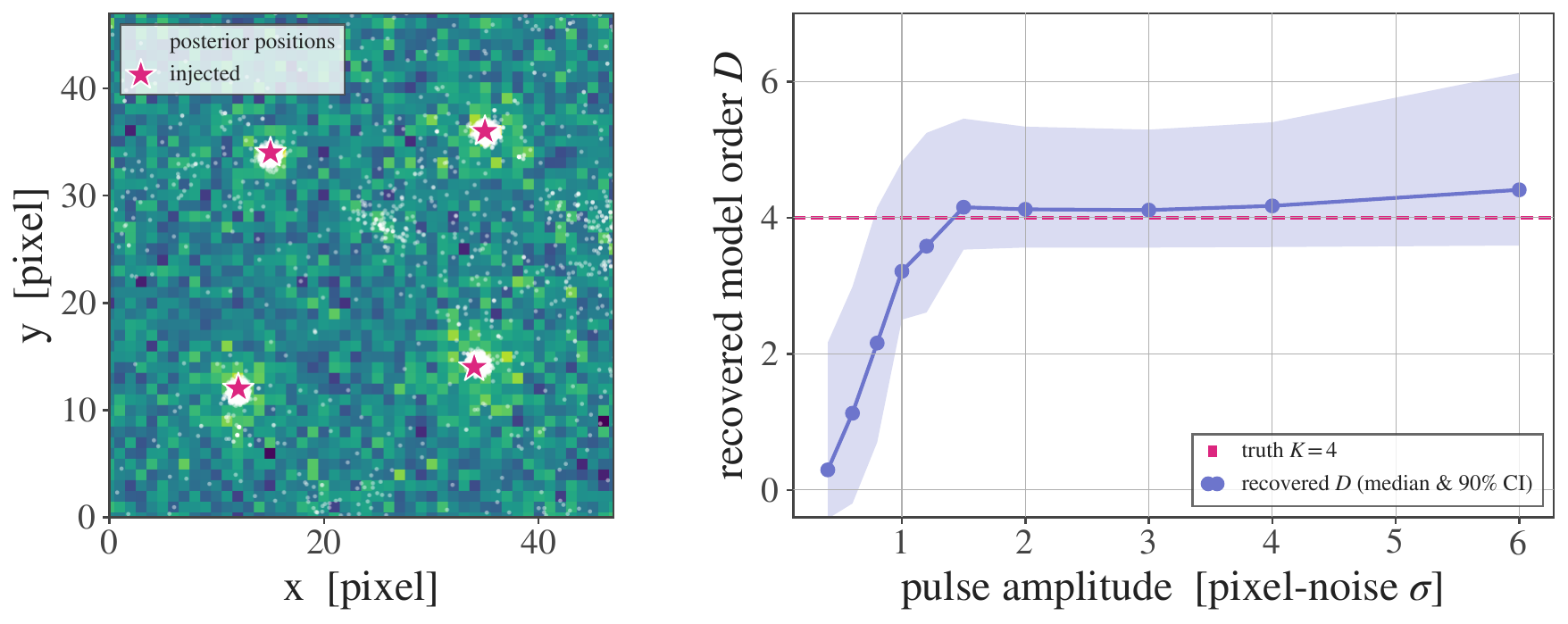}
\caption{Model selection as generic detection: an unknown number of 2-D Gaussian pulses in pixel noise (left: the image at pulse amplitude $2\sigma$, with the injected positions and the posterior position samples). The recovered model order (right) climbs from the empty model to the true $K=4$ as the pulses cross the detection threshold and locks on --- the image-domain counterpart of the wavelet experiment, with no domain-specific structure. Quantiles of the discrete $D$ use the continuity correction of Methods~\ref{sec:converge}.}
\label{fig:pulses2d}
\end{figure}

\begin{figure}[!tbp]
\centering
\includegraphics[width=0.6\columnwidth]{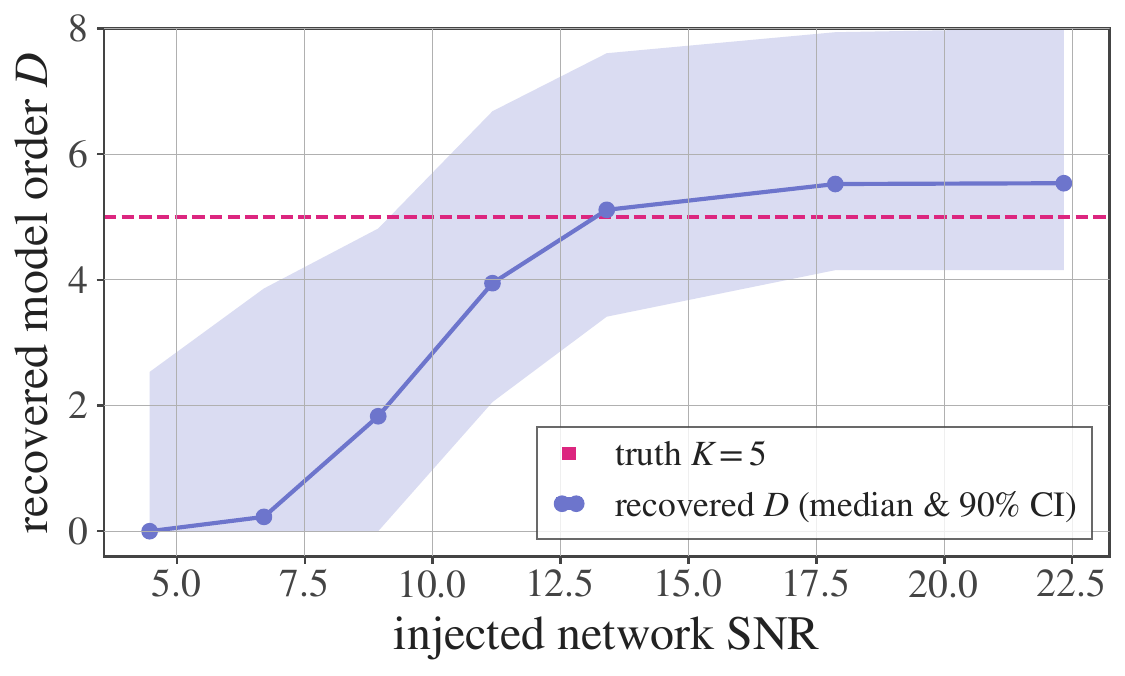}
\caption{Model selection on a known injection ($K=5$ wavelets in white noise). The recovered model order (median and $90\%$ credible band) climbs from the empty model to the truth as the wavelets become detectable and locks on. The companion posterior $p(D\,|\,{\rm data})$ at network SNR $13$ peaks exactly at $K=5$. Quantiles of the discrete $D$ use the continuity correction of Methods~\ref{sec:converge}.}
\label{fig:modelsel}
\end{figure}

Because the number of components is itself inferred, we can test whether $p(D\,|\,{\rm data})$ recovers a known truth --- and the cleanest such test requires no domain background at all. An unknown number of 2-D Gaussian pulses is injected into pixel noise --- the abstraction of source counting in astronomical images, or of feature counting in microscopy and medical imaging --- and the sampler, with no image-specific structure whatsoever, must count them (Fig.~\ref{fig:pulses2d}). As the pulse amplitude is swept upward the posterior traces the detection transition textbook-style: it favours the empty model when the pulses are undetectable, climbs as they cross the threshold, and locks onto the true $K=4$ once each pulse exceeds $\sim\!1.5\times$ the pixel noise. The same experiment in the time--frequency domain gives the same answer: $K=5$ well-separated Morlet--Gabor wavelets injected into white noise at a controlled per-wavelet SNR $\rho$ (Fig.~\ref{fig:modelsel}); once all five are detectable (network SNR $\gtrsim 13$), $p(D\,|\,{\rm data})$ concentrates on the true $K=5$: at network SNR $13.4$ the posterior is $p(4){\approx}0.06$, $p(5){\approx}0.33$--$0.37$ across seeds, $p(6){\approx}0.31$, $p(7){\approx}0.18$ --- a clear mode at the truth with the mass spread over its neighbours rather than a sharp spike. In both settings the recovered order saturates near the truth with a mild $+1$ tendency at very high SNR. This is not an Occam effect --- the marginal likelihood and the $p(D)$ prior both penalise extra components --- but a consequence of the amplitude prior of Eq.~\eqref{eq:rhoprior}, which retains a small but non-zero mass at low $\rho$ and therefore permits an occasional spare, low-amplitude component to absorb a noise fluctuation at negligible likelihood cost. These are direct, ground-truthed demonstrations of the trans-dimensional capability underlying every reconstruction in this paper.

\begin{figure}[!tbp]
\centering
\includegraphics[width=0.9\linewidth]{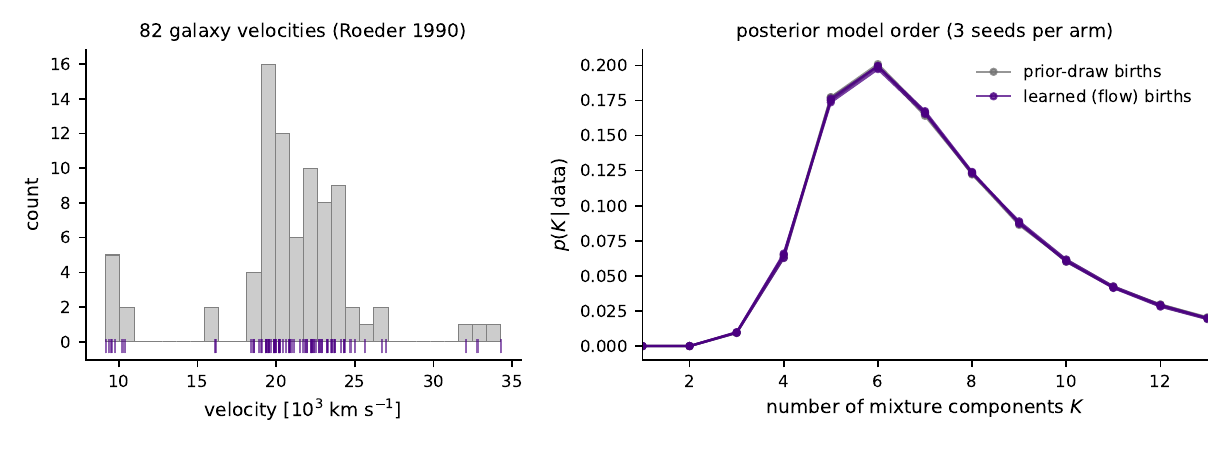}
\caption{The canonical trans-dimensional benchmark: a Gaussian mixture with unknown $K$ on the $82$ galaxy velocities~\cite{Roeder1990,Richardson1997}. Left: the data. Right: $p(K\,|\,\mathrm{data})$ for six independent runs --- three seeds with prior-draw births (grey) and three with learned births (indigo) --- which coincide to within line width; the posterior mode is $K=6$ with support on $4$--$9$, consistent with published analyses of these data. $\hat\tau_K\approx3$: model-order mixing is not a bottleneck here, and the learned proposal neither helps nor hurts, as the phase argument predicts.}
\label{fig:galaxy}
\end{figure}

Finally, we place the sampler on the canonical benchmark of the trans-dimensional literature itself: the $82$ galaxy recession velocities~\cite{Roeder1990} on which Richardson \& Green introduced trans-dimensional mixture estimation~\cite{Richardson1997}. A Gaussian mixture with unknown $K$ (uniform priors on the component means, log-widths and unnormalised weights; $K\le15$; Methods~\ref{sec:interface} defaults) yields a posterior concentrated on $K=4$--$9$ with mode $K=6$ --- $p(6)\approx0.20$, $p(5)\approx0.18$, $p(7)\approx0.17$ --- stable to three decimals across six runs (three seeds for each of two birth proposals; Fig.~\ref{fig:galaxy}), in the range reported across published analyses of these data, whose $p(K)$ is well known to be sensitive to the component priors~\cite{Richardson1997}. On this small problem the model-order chain decorrelates in a few steps ($\hat\tau_K\approx3$) and learned births leave the posterior unchanged --- as the phase argument itself predicts, a learned proposal has nothing to add where model-order mixing is not the bottleneck. The value of the experiment is external: a like-for-like validation of the identical code path on the problem the trans-dimensional literature knows best.

\subsection{Generality across domains and instruments}
\label{sec:showcase}

\begin{figure}[!tbp]
\centering
\begin{subfigure}{0.73\linewidth}\centering
  \includegraphics[width=\linewidth]{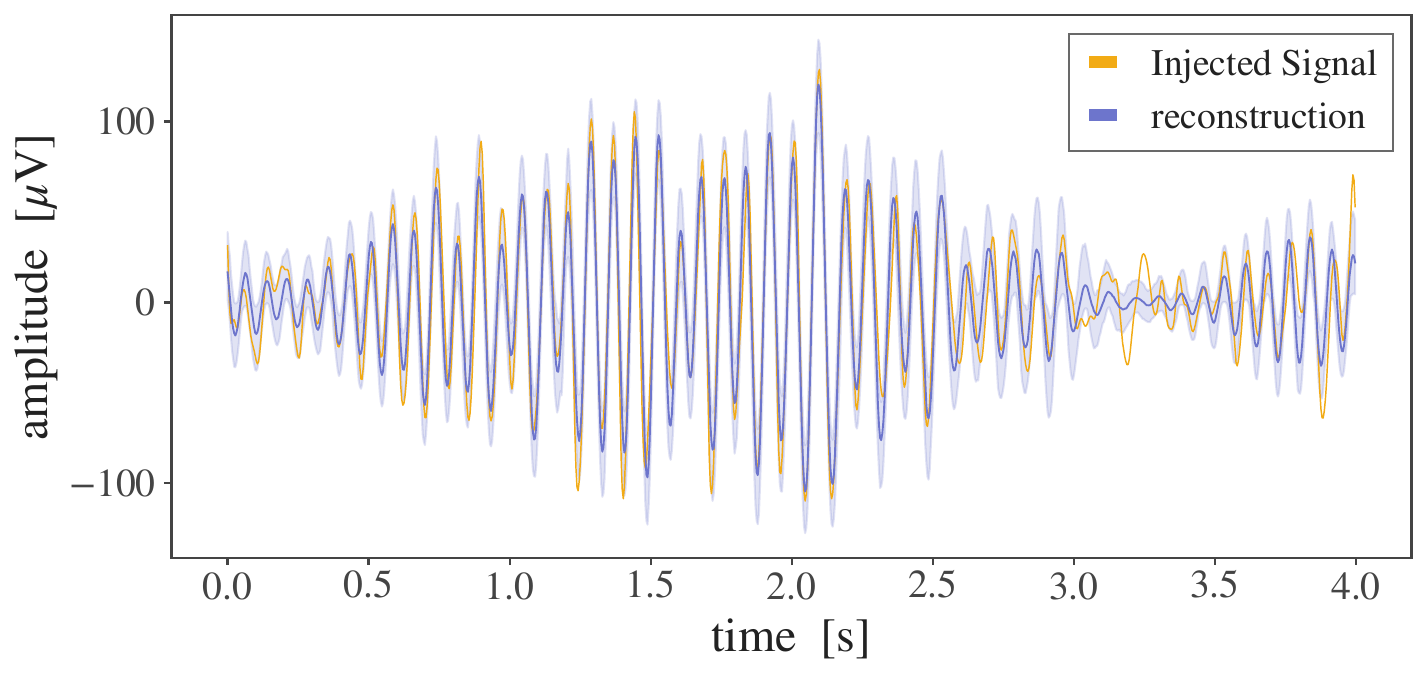}
  \caption{Human EEG, CHB-MIT (time domain).}
  \label{fig:sc-ecg}
\end{subfigure}\\[4pt]
\begin{subfigure}{0.73\linewidth}\centering
  \includegraphics[width=\linewidth]{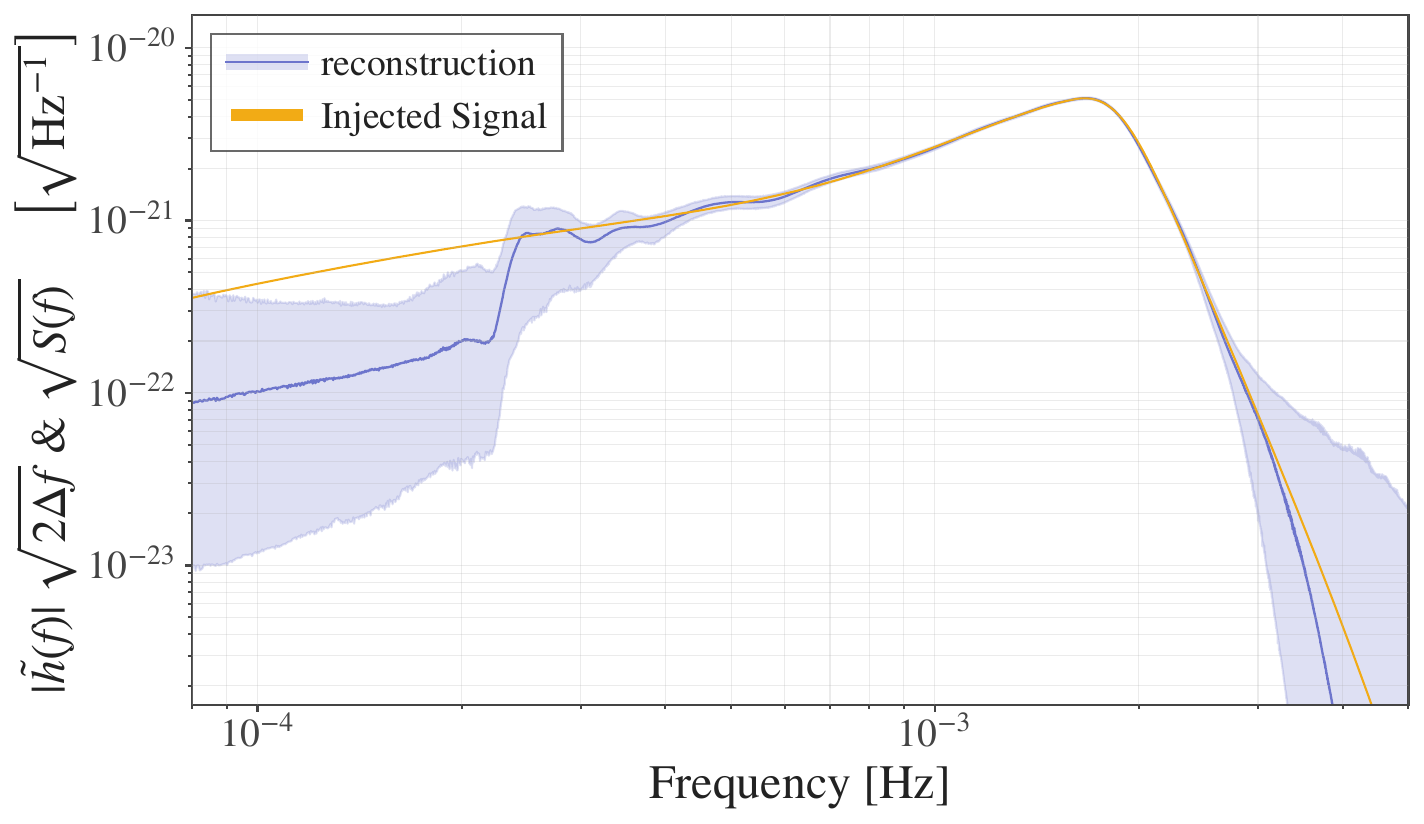}
  \caption{LISA massive black-hole binary, \textsc{Sangria} (frequency domain).}
  \label{fig:sc-lisa}
\end{subfigure}\\[4pt]
\begin{subfigure}{0.73\linewidth}\centering
  \includegraphics[width=\linewidth]{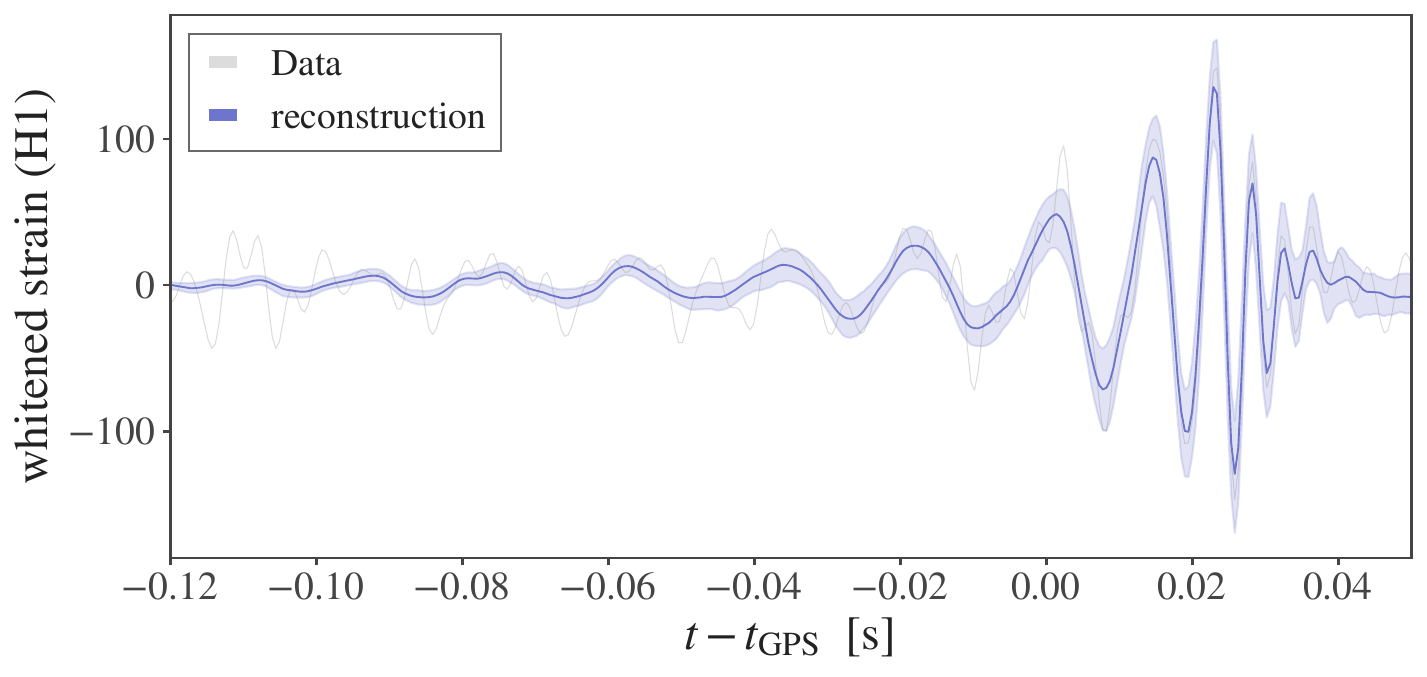}
  \caption{GW150914, real LIGO data (time domain).}
  \label{fig:sc-gw}
\end{subfigure}
\caption{One sampler across two domains and six orders of magnitude in frequency. Model-agnostic wavelet reconstruction (indigo: posterior median and $90\%$ credible band) against the reference signal: (a,\,b) the injected signal (orange) for the band-limited clean EEG and the simulated LISA binary; (c) whitened GW150914 strain --- no injected truth, so the whitened data is shown. The same reversible-jump sampler and code path produce all three.}
\label{fig:showcase}
\end{figure}

The reconstruction model makes no assumption specific to a given instrument or even to physics: any whitened, band-limited series is reconstructed by supplying a linear response operator (a unit response recovers the series itself). We illustrate this in order of increasing domain complexity --- a human physiological signal, a simulated space-based detection, and finally real interferometer data --- spanning frequency bands separated by six orders of magnitude (Fig.~\ref{fig:showcase}, Table~\ref{tab:showcase}).

\emph{A human brain rhythm.} A four-second scalp-EEG excerpt (CHB-MIT database, channel FP2--F4; a transient $\sim$12~Hz oscillation of the kind central to sleep and seizure neurophysiology), buried in white noise at matched-filter SNR $25$, is recovered to a match of $0.883$ against the clean trace with a median of $9$ wavelets (Fig.~\ref{fig:showcase}a) --- with a posterior credible band, and with the number of components inferred rather than assumed.

\emph{A space-based gravitational-wave source.} A simulated LISA massive black-hole binary in the mHz band (\textsc{Sangria} data set~\cite{LDC}, noise-orthogonal TDI channel, SNR $351$) reaches a network match of $0.999$ with a median of $17$ wavelets --- the same code, six orders of magnitude away in frequency and on a different instrument response.

\emph{Real interferometer data: GW150914.} The heaviest test is public LIGO Hanford/Livingston strain around GPS $1126259462.4$ (event GW150914~\cite{Abbott2016,Abbott2019gwtc1}), where the black-hole merger is recovered model-agnostically after automatic data conditioning (Methods~\ref{sec:conditioning}) removes instrumental lines and sets the analysis band. The reconstruction recovers a network SNR of $24.9$ (consistent with the LVC matched-filter value $\approx24$) with a whitened-residual $\chi^2/\mathrm{dof}=0.99$ --- the residual is statistically pure noise, i.e.\ the model is neither over- nor under-fitting --- at a median of $D\!\sim\!18$ wavelets. Against the best-fit compact-binary template (GWTC-1 parameters~\cite{Abbott2019gwtc1}), the median reconstruction attains a phase- and time-maximised \emph{network} match of $0.97$ ($0.98$ in H1), confirming faithful extraction of the astrophysical signal. This is comparable to the fidelity reported for the reference \textsc{BayesWave} analysis of the same event ($\sim\!0.94$~\cite{Cornish2021,Ghonge2020}); we stress that the two numbers are not a controlled comparison, since the published \textsc{BayesWave} match is computed against the numerical-relativity waveform under that pipeline's own data conditioning, whereas ours is phase- and time-maximised against the GWTC-1 template under the conditioning of Methods~\ref{sec:conditioning}. Maximisation alone raises a match, and we make no claim of superiority on this basis; the controlled comparisons in this paper are the within-code ones of Table~\ref{tab:bench}. The sampled extrinsic branch --- the sky position, polarization and ellipticity shared by all wavelets (Methods~\ref{sec:notation}) --- simultaneously localizes the source: the (RA, Dec) posterior forms the characteristic two-detector arc in the southern sky with a $90\%$ credible area of $142$~deg$^2$ ($50\%$: $31$~deg$^2$), consistent with the published localization of the event~\cite{Abbott2019gwtc1}.

\begin{table}[t]
\caption{One sampler, two domains and six orders of magnitude of frequency: model-agnostic reconstruction fidelity (network match to the reference, Eq.~\eqref{eq:match}).}
\label{tab:showcase}
\centering
\begin{tabular}{lcccc}
\toprule
Signal & SNR & Match & $D$ & Wall \\
\midrule
Human EEG (CHB-MIT)   & $25$   & $0.883$ & $9$  & $110$~min \\
LISA MBHB (Sangria)   & $351$  & $0.999$ & $17$ & $79$~min \\
GW150914 (LIGO data)  & $24.9$ & $0.97$  & $18$ & $32$~min \\
\bottomrule
\end{tabular}
\\[2pt]
{\footnotesize Wall-clock to satisfaction of the stopping rule (Methods~\ref{sec:converge}) on a single NVIDIA A100, measured before the environment change of Methods~\ref{sec:repro}. Wall times are not comparable across rows: the three analyses differ in data length, band, and $D_{\max}$.}
\end{table}

\subsection{Computational cost}
\label{sec:cost}

On a single GPU the GW150914 reconstruction satisfies its stopping rule within a working session (Table~\ref{tab:showcase}), recovering the event at a network SNR of $24.9$ with a whitened-residual $\chi^2/\mathrm{dof}=0.99$, and matching the best-fit compact-binary template at $0.97$; on the controlled benchmark of Table~\ref{tab:bench} the flow-assisted reconstruction reaches an overlap of $0.954$. Re-run end-to-end on the current computing environment (Methods~\ref{sec:repro}), the same GW150914 analysis completes in $39$~min with statistically identical results (network SNR $25.1$, $\chi^2/\mathrm{dof}=0.98$, $D\sim18$) --- the reconstruction quality is insensitive to the environment change that rescaled the benchmark's step counts. The reference \textsc{BayesWave} RJMCMC~\cite{Cornish2015,Cornish2021} requires hours of runtime on multi-core CPUs at comparable fidelity, but we do not present this as a controlled speed-up: the two analyses differ in hardware, data conditioning and stopping criteria simultaneously. The controlled statement is the within-code one: on identical hardware, with identical data and identical stopping rule, the Fisher-only cascade fails the rule within twice the wall time of the flow-assisted cascade (Table~\ref{tab:bench}), so the acceleration is attributable to the learned proposal and not to the GPU alone.

The framework is also competitive for fixed-dimensional, template-based inference: with a heterodyned (relative-binning) likelihood it reproduces a \textsc{bilby}--\textsc{dynesty} posterior for a $64$~s binary-black-hole signal in Gaussian noise to a maximum per-parameter Jensen--Shannon divergence of $0.006$ (statistically indistinguishable) in $11$~min against $73$~min. This is a same-hardware comparison --- both pipelines ran in the same batch job on the same $16$-core AMD CPU node with no GPU --- but it is a \emph{pipeline}-level comparison, not a sampler-level one: our run uses the heterodyned likelihood, whereas the \textsc{bilby} reference uses the standard full likelihood, and the speed-up is attributable to that fast likelihood path. Run with the identical full Gaussian likelihood, the ensemble sampler is in fact \emph{slower} than \textsc{dynesty} on this problem ($169$ against $73$~min). The claim is therefore that the packaged pipeline reproduces the reference posterior at a fraction of the cost out of the box --- not that the sampler is faster at equal likelihood cost.

\subsection{Exactness, and when the gain materialises}
\label{sec:fixed-pe}

The learned proposal is not specific to the trans-dimensional problem, but its benefit is confined to settings where the move it feeds is the bottleneck. Applied to ordinary $11$-parameter estimation of the \textsc{Sangria} massive black-hole binary with \textsc{Eryn}, a flow independence proposal (added under an exact Metropolis--Hastings correction, so it cannot bias the posterior) raises the in-model acceptance by $46\%$ ($0.085$ to $0.124$, three seeds). Because the parallel-tempered ensemble already explores this particular posterior efficiently --- its integrated autocorrelation time is only a few steps --- the higher acceptance does not translate into a material effective-sample-size gain here. The exactness guarantee is precisely what makes the proposal safe to attempt regardless: the Metropolis--Hastings correction leaves the target invariant for any flow, so the proposal can always be added and contributes exactly where the kernel it feeds limits performance --- in the trans-dimensional regime above, the model-order mixing driven by the birth moves. Flow transport maps are also used to great effect for sampling and evidence estimation within preconditioned sequential Monte Carlo~\cite{Karamanis2022}, a complementary approach with different aims.

\section{Discussion}
\label{sec:conclusions}

We have shown that the value of a learned proposal in trans-dimensional inference is decided by a phase structure (Fig.~\ref{fig:phases}): the optimal birth is residual-matched during assembly and degenerates to the posterior's component marginal at equilibrium --- the object an adaptive flow, trained online and corrected by exact Metropolis--Hastings, learns for free. Exploiting this division of labor brings a pre-specified stopping point within reach several times sooner, without altering the target distribution --- a point made precise by the structural argument and the controlled, isolation, and ablation experiments of Sec.~\ref{sec:results-births}. Two features make the result broadly useful. First, the guarantee of exactness means the network may be retrained aggressively and used without validation of its accuracy, in contrast to methods that bake a learned map into the target. Second, the accelerator is agnostic to the application: the same software reconstructs a black-hole merger recorded by kHz ground-based detectors, a mHz signal from a future space mission, and a human electroencephalogram, counts sources in a noisy image, and reproduces reference template-based posteriors at a fraction of the cost.

The natural extension follows directly from the phase structure we identify. A flow conditioned on a low-dimensional summary of the current whitened residual --- a coarse time--frequency map, or the leading $\mathcal{F}$-statistic peaks --- would interpolate between the two regimes by construction: residual-matched during assembly, population-matched at equilibrium. It could in principle be trained to approximate $g^\star$ directly, within the reversible-jump~\cite{Green1995} or involutive~\cite{Neklyudov2020} frameworks that guarantee validity. Whether such a conditional proposal can be trained online cheaply enough to beat the closed-form residual matched filter it would replace is an open and, we think, worthwhile question.

Within gravitational-wave astronomy alone the mechanism meets several open needs. Model-agnostic burst reconstruction --- the regime \textsc{BayesWave} was built for~\cite{Cornish2015,Cornish2021} --- inherits the same stopping behaviour on single-GPU timescales. Post-merger remnants of binary neutron-star mergers, whose waveform is uncertain and whose detectability with next-generation detectors we quantified previously~\cite{Sasli2024}, are precisely the short, unmodeled transients the wavelet model targets. Because the shared sky position is inferred jointly with the reconstruction, each such analysis delivers a localization posterior as a byproduct (Sec.~\ref{sec:showcase}) --- for sources with no low-latency template-based counterpart, this is the localization that electromagnetic follow-up would otherwise lack. And in the reversible-jump global-fit pipelines of LISA~\cite{Littenberg2020,Katz2024}, where tens of thousands of galactic binaries are resolved and refined jointly, learned population-matched birth proposals accelerate exactly the model-order mixing that dominates the refinement phase --- while, as our structural argument shows, residual-matched births remain the right tool for \emph{discovering} sources in the first place.

Trans-dimensional model selection --- counting change points, sources in images or data streams, mixture components, or spectral features --- is ubiquitous across the physical, life, and data sciences; because the sampler requires only a likelihood and a component template, the mechanism demonstrated here transfers directly, and we provide it as an off-the-shelf open-source tool. The software, examples, and data to reproduce every figure are publicly available.

\section*{Methods}
\setcounter{section}{4}
\setcounter{subsection}{0}

\subsection{Notation, definitions, and reported quantities}
\label{sec:notation}

All inner products are noise-weighted. For two frequency-domain series $a,b$ in detector $I$ with one-sided power spectral density $S_I(f)$,
\begin{equation}
\label{eq:inner}
\langle a\,|\,b\rangle_I = 4\,\mathrm{Re}\!\!\int_{f_{\rm low}}^{f_{\rm high}}\!\!
   \frac{\tilde a(f)\,\tilde b^{*}(f)}{S_I(f)}\,\mathrm{d}f ,
\qquad
\langle a\,|\,b\rangle = \sum_I \langle a\,|\,b\rangle_I ,
\end{equation}
the latter being the network inner product. The following quantities are used throughout.

\paragraph{Signal-to-noise ratio.} The optimal network SNR of a waveform $h$ is $\rho_{\rm opt}=\sqrt{\langle h|h\rangle}$ and the matched-filter SNR in data $d$ is $\rho_{\rm mf}=\langle d|h\rangle/\sqrt{\langle h|h\rangle}$. Injected SNRs quoted for simulations are optimal SNRs of the injected signal; the SNR quoted for a reconstruction (e.g.\ $24.9$ for GW150914) is $\sqrt{\langle h_{\rm med}|h_{\rm med}\rangle}$ evaluated on the posterior-median reconstruction $h_{\rm med}$. The per-wavelet SNR $\rho$ is the optimal SNR induced by a single wavelet in isolation and is the amplitude parameter of the model (Sec.~\ref{sec:model}).

\paragraph{Overlap, or match.} We use the two words interchangeably for the normalised, phase- and time-maximised network inner product between a reconstruction $h$ and a reference waveform $s$,
\begin{equation}
\label{eq:match}
\mathcal{M}(h,s) = \max_{t_c,\phi_c}
   \frac{\langle h\,|\,s\rangle}{\sqrt{\langle h|h\rangle\,\langle s|s\rangle}} \in [-1,1].
\end{equation}
Where a single-detector value is quoted (e.g.\ ``$0.98$ in H1'') the sums in Eq.~\eqref{eq:inner} are restricted to that detector. Maximisation over $(t_c,\phi_c)$ raises $\mathcal{M}$ relative to an unmaximised overlap; matches taken from the literature are not necessarily maximised in the same way, and we therefore compare maximised matches only within this work.

\paragraph{Wavelet, or atom.} The elementary component of the model is a single Morlet--Gabor wavelet, Eq.~\eqref{eq:wavelet}. We refer to it as a wavelet throughout; where the model order $D$ is described as a number of ``atoms'' the two terms are synonymous.

\paragraph{Intrinsic and extrinsic parameters.} The full parameter vector is $\theta=\{D,\{t_0,f_0,Q,\rho,\phi_0\}_{1:D},\Omega\}$. The per-wavelet parameters are \emph{intrinsic}; the block $\Omega=(\alpha,\delta,\psi,\epsilon)$ --- sky position, polarization angle and ellipticity --- is shared by all $D$ wavelets and is called the \emph{extrinsic branch}. It is sampled jointly with the rest, which is why a sky localization is obtained as a byproduct of the reconstruction.

\paragraph{Acceptance rates.} \emph{In-model acceptance} is the acceptance fraction of within-model (fixed-$D$) Metropolis--Hastings moves on the cold chain, averaged over the post-burn-in samples. \emph{Reversible-jump acceptance} is the acceptance fraction of birth and death moves, reported separately. The two are not comparable and are never combined.

\paragraph{Ensemble.} The sampler runs $N_{\rm temp}$ parallel-tempered chains (temperature ladder $\beta_1{=}1 > \beta_2 > \dots$), each an ensemble of $N_{\rm walk}$ walkers. Unless stated otherwise we use $N_{\rm temp}=10$ with the adaptive ladder of \textsc{Eryn} (temperatures adjusted on the fly toward uniform swap acceptance, $T_{\max}=\infty$) and $N_{\rm walk}=50$. All diagnostics ($\hat\tau$, ESS, acceptance rates, $p(D)$) are computed on the cold chain $\beta=1$ only.

\subsection{Wavelet model, priors, and likelihood}
\label{sec:model}

Each Morlet--Gabor wavelet is a sine-Gaussian localized at time $t_0$ and frequency $f_0$ with quality factor $Q$, amplitude $A$, and phase $\phi_0$. Its frequency-domain plus polarization, written for $f>0$, is
\begin{equation}
\label{eq:wavelet}
\Psi(f) = \frac{A\,\tau\sqrt{\pi}}{2}\,e^{-2\pi i f t_0}
  \Big[ e^{i\phi_0} e^{-\pi^2\tau^2 (f-f_0)^2}
      + e^{-i\phi_0} e^{-\pi^2\tau^2 (f+f_0)^2} \Big],
\end{equation}
with $\tau = Q/(2\pi f_0)$. We reparameterize the amplitude by the induced single-wavelet optimal signal-to-noise ratio $\rho$, $A\!\propto\!\rho$, and place a peaked prior
\begin{equation}
\label{eq:rhoprior}
p(\rho)\propto \rho\,\big[1+(\rho/2\rho_\star)^2\big]^{-1}
\end{equation}
that suppresses spurious low-amplitude wavelets~\cite{Cornish2015}, with $\rho_\star=5$ for the benchmark and the pulse experiments, $3$ for GW150914, $10$ for the LISA binary, and $6$ for the EEG. The remaining priors are: $t_0$ uniform over the analysis segment; $f_0$ uniform over the analysis band ($[20,512]$~Hz benchmark, $[32,350]$~Hz GW150914, $[0.1,4]$~mHz LISA, $[4,40]$~Hz EEG); $Q$ uniform on $[0.1,40]$; $\phi_0$ uniform on $[0,2\pi)$; $\Omega$ isotropic on the sphere with $\psi$ uniform on $[0,\pi)$ and $\epsilon$ uniform on $[-1,1]$; and $p(D)$ uniform on $[0,D_{\max}]$ with $D_{\max}=40$ (benchmark), $25$ (GW150914), $20$ (LISA), $20$ (EEG). A source is a set of $D$ wavelets sharing the extrinsic state $\Omega$ (Sec.~\ref{sec:notation}). The strain in detector $I$ is
\begin{equation}
\label{eq:project}
h_I(f;\theta) = \big[F^+_I(\Omega)\,\Psi^\Sigma(f) + i\,\epsilon\,F^\times_I(\Omega)\,\Psi^\Sigma(f)\big]
                e^{-2\pi i f\,\Delta t_I(\Omega)},
\end{equation}
where $\Psi^\Sigma=\sum_{j=1}^{D}\Psi_j$, $F^{+,\times}_I$ are the antenna responses of detector $I$ (Advanced LIGO~\cite{Aasi2015}, Advanced Virgo~\cite{Acernese2015}) and $\Delta t_I$ the geocentric time delay; setting $F^+_I\!\equiv\!1,F^\times_I\!\equiv\!0,\Delta t_I\!\equiv\!0$ recovers the whitened series itself, which is what makes the model instrument-agnostic (Sec.~\ref{sec:showcase}). Assuming stationary Gaussian noise with one-sided power spectral density $S_I(f)$, the Whittle log-likelihood is
\begin{equation}
\label{eq:like}
\ln\mathcal{L}(\theta) = -2\Delta f\sum_{I}\sum_{k}
   \frac{\big|d_I(f_k)-h_I(f_k;\theta)\big|^2}{S_I(f_k)} + \text{const}.
\end{equation}
Every wavelet across all $N_{\rm temp}\!\times\!N_{\rm walk}$ walkers is generated and projected by a single vectorized kernel dispatched to \texttt{NumPy} (CPU) or \texttt{CuPy} (GPU); the batched device path removes the per-walker loop and is what makes tempered ensembles with tens of wavelets affordable.

\subsection{Automatic data conditioning for real strain}
\label{sec:conditioning}

Real interferometer data are neither stationary nor featureless, and a model-agnostic wavelet basis will faithfully reconstruct instrumental artefacts unless they are removed first. \hyperwave{} therefore applies an automatic conditioning stage before sampling, with no per-event hand tuning.

\begin{enumerate}
\item \emph{Segment selection and windowing.} For GW150914 a $4$~s segment centred near the GPS trigger is extracted at $2048$~Hz and tapered with a Tukey window (roll-off fraction $0.2$), chosen so that the $f<20$~Hz seismic wall does not leak into the analysis band.
\item \emph{PSD estimation.} The one-sided PSD $S_I(f)$ is estimated by median-Welch averaging over $512$~s of off-source data, interpolated to the analysis resolution.
\item \emph{Band selection.} An automatic low-cut is available: the whitened per-bin power $w(f)=4\Delta f\,|d|^2/S$ is smoothed with a $40$-bin running mean, and $f_{\rm low}$ is placed at the lowest frequency where the smoothed floor drops below $4$ and stays below $8$ --- a running \emph{mean} rather than median, so that a dense low-frequency line cluster keeps the floor high until the band genuinely clears it. For GW150914 the automatic cut lands at $48$~Hz, which sacrifices the $35$--$48$~Hz early inspiral ($16\%$ of the template signal power); we therefore fix the band to $[32,350]$~Hz and rely on the line notching below, which retains $92\%$ of the signal power.
\item \emph{Line identification and soft notching.} A bin is flagged as a line if its whitened power exceeds the running median by $6\times1.4826\,\mathrm{MAD}$ (window $12$ bins) over a contiguous run of at most $4$ bins --- lines are narrow, broadband features are not --- with a protected band ($[120,260]$~Hz, containing the merger) never flagged. Flagged bins (plus the explicitly listed $60$~Hz mains harmonics and the $35$--$40$~Hz suspension cluster) are \emph{soft}-notched: their PSD is inflated by a factor $10^{20}$, driving their likelihood weight to zero without removing bins, so template and data keep identical shapes. Left in place, these lines carry roughly half of the apparent excess power and inflate the recovered network SNR from $\approx25$ to $\approx44$.
\item \emph{Whitening and self-check.} The conditioned data are whitened by $\sqrt{S_I(f)}$, and a whiteness report measures the whitened power in probe bands (expected value $2$ per bin for pure noise); the stage \emph{warns} if the noise floor departs from $\sim2$ --- a stray $\sqrt{2}$ in the PSD or window normalization lands it at $1$ or $4$ --- or if strong lines remain. For GW150914 the post-conditioning floor is $2.00$ and the residual $\chi^2/\mathrm{dof}$ of Sec.~\ref{sec:showcase} is referred to unit-variance noise.
\end{enumerate}

The same stage is a no-op for simulated data with known PSD (LISA, the toy injections) and reduces to whitening alone for the EEG application.

\subsection{Reversible-jump sampler}
\label{sec:rjmcmc}
We sample the joint posterior $\pi(\theta)\propto\mathcal{L}(\theta)\,p(\theta)$ with the parallel-tempered RJMCMC of \textsc{Eryn}~\cite{Karnesis2023}. Note that $\pi$ is fixed by the data and the priors; it is the distribution of the chain that evolves towards $\pi$, and the adaptation described in Sec.~\ref{sec:flowprop} changes only how the chain explores $\pi$, never $\pi$ itself.

A within-model move at fixed $D$ proposing $\theta\!\to\!\theta'$ from a proposal kernel $q(\theta'|\theta)$ is accepted with probability
\begin{equation}
\label{eq:mh}
\alpha = \min\!\left[1,\;
   \frac{\pi(\theta')}{\pi(\theta)}\,
   \frac{q(\theta\,|\,\theta')}{q(\theta'\,|\,\theta)}\right].
\end{equation}
Here $q$ is any valid within-model kernel; our default is the cascade described at the end of this subsection, and the learned variant $q_\varphi$ is introduced in Sec.~\ref{sec:flowprop}.

A birth move $D\!\to\!D\!+\!1$ draws the new wavelet's parameters $u\in\mathbb{R}^5$ from a \emph{birth proposal} $g(u\,|\,\theta)$, a different object from $q$: $q$ perturbs existing wavelets at fixed $D$, whereas $g$ generates the parameters of one additional wavelet. With the identity dimension-matching map the birth acceptance is~\cite{Green1995}
\begin{equation}
\label{eq:rj}
\alpha_{\rm birth} = \min\!\left[1,\;
   \frac{\mathcal{L}(\theta')}{\mathcal{L}(\theta)}\,
   \frac{p(D\!+\!1)}{p(D)}\,\frac{p(u)}{(D\!+\!1)\,g(u\,|\,\theta)}\,
   \frac{d_{D+1}}{b_D}\right],
\end{equation}
with $b_D,d_{D+1}$ the birth/death selection probabilities ($b_D=d_{D+1}=0.5$ except at the boundaries, the \textsc{Eryn} default); deaths use the reciprocal ratio. Our default $g$ is data-informed: a matched-filter fit of a single wavelet to the whitened residual $r=d-h(\theta)$ sets its amplitude and phase in closed form, with $t_0$ and $f_0$ drawn from the whitened data's marginal time- and frequency-power profiles (each tabulated on a $512$-point grid and floored at $10\%$ of its peak so the full band retains support), and $Q$ from its prior. In-model updates use a cascade of a Fisher-matrix Gaussian move, a half-cycle move that jointly shifts $(t_0,\phi_0)$ by half a period (the classic \textsc{BayesWave} degeneracy trick), and a sky-ring move on $\Omega$, mixed with weights $0.7$ (Fisher), $0.1$ (half-cycle), $0.1$ (sky-ring), and $0.1$ for a broad Gaussian fallback kernel.

\subsection{Adaptive normalizing-flow proposal}
\label{sec:flowprop}
The flow variant adds an independence proposal $q_\varphi(\theta')$ that does \emph{not} depend on the current state. We use a neural spline flow~\cite{Durkan2019} --- from the normalizing-flow density-estimation lineage~\cite{Rezende2015,Papamakarios2017} --- $T_\varphi:\mathbb{R}^5\!\to\!\mathbb{R}^5$ mapping a base density $p_0$, taken to be a standard multivariate normal $\mathcal{N}(0,\mathbb{I}_5)$, to the single-wavelet parameter space (with the periodic phase mapped to the circle); its density follows from the change of variables,
\begin{equation}
\label{eq:flow}
\log q_\varphi(\theta') = \log p_0\!\big(T_\varphi^{-1}(\theta')\big)
   + \log\Big|\det \frac{\partial T_\varphi^{-1}}{\partial\theta'}\Big|.
\end{equation}
Because the proposal exposes both a sampler ($\theta'=T_\varphi(z),z\sim p_0$) and the exact density~\eqref{eq:flow}, it enters the independence-sampler form of~\eqref{eq:mh}, $q(\theta'|\theta)=q_\varphi(\theta')$, and the Metropolis--Hastings ratio corrects for \emph{any} imperfection of the flow: the invariant distribution of the chain remains exactly $\pi$. Because this holds for \emph{any} flow, the network can be retrained aggressively without biasing the result and without validating its accuracy against the target. This is the same guarantee exploited by the adaptive scheme of Ref.~\cite{Gabrie2022pnas}, its implementation \textsc{flowMC}~\cite{Wong2023flowMC}, and \textsc{jim}~\cite{Wong2023jim} at fixed dimension.

\paragraph{Architecture and training.} The flow is the \texttt{pocomc} \texttt{nsf3} preset over \texttt{zuko}: three masked-autoregressive transforms with monotonic rational-quadratic-spline bases ($8$ bins), conditioner networks of width $32$ with residual blocks and ReLU activations. Before training, $t_0,f_0,Q,\rho$ are affinely rescaled to the prior box and the periodic phase $\phi_0$ is handled on the circle. The parameters $\varphi$ are fit by maximizing the likelihood of the cold-chain samples $\{\theta_i\}\sim\pi$ accumulated since the last update,
\begin{equation}
\label{eq:train}
\varphi^\star = \arg\max_\varphi \sum_i \log q_\varphi(\theta_i)
   \;\equiv\; \arg\min_\varphi \mathrm{KL}\!\big(\pi \,\|\, q_\varphi\big),
\end{equation}
i.e.\ standard forward-KL density estimation. The flow is retrained every $N_{\rm up}=100$ iterations from the cold chain and falls back to the prior until at least $256$ samples are available; $\varphi$ is warm-started from the previous update. Three implementation details matter for validity. (i)~The training set pools the active wavelets of all cold-chain walkers accumulated since the last update, so a state with $D$ wavelets contributes $D$ samples; the learned density is therefore the occupancy-weighted component marginal --- precisely the object the equilibrium argument of Methods~\ref{sec:births} concerns. (ii)~The birth proposal is a \emph{defensive mixture}: the flow is mixed with the prior at weight $0.15$, so the kernel retains support on the full prior box and irreducibility never rests on the network. (iii)~The flow is not frozen, so the transition kernel is time-inhomogeneous; each proposal is applied with its exact density in the acceptance ratio at the moment it is used, adaptation diminishes naturally as the training history grows, and all reported diagnostics are computed on the final chain. The retraining cost is fully absorbed on the GPU: in the benchmark of Table~\ref{tab:bench} the flow-assisted configuration averaged $0.199$~s per iteration \emph{including} retraining, against $0.234$~s for the Fisher-only cascade, so the reported wall times are net of all training overhead. In the benchmark the trained flow serves as the \emph{birth-move} generator of Eq.~\eqref{eq:rj} (Sec.~\ref{sec:results-births}); when used instead as an in-model independence move it is mixed into the cascade with weight $w_{\rm flow}=0.3$.

\subsection{Two phases, two optimal birth proposals}
\label{sec:births}
Reusing $q_\varphi$ as the birth proposal $g$ in~\eqref{eq:rj} leaves the reversible-jump \emph{acceptance} unchanged ($0.09$--$0.10$ for flow births against $0.09$--$0.11$ for the data-informed $g$, both in the isolation experiments and in Table~\ref{tab:bench}), yet it is the configuration that meets the stopping rule. The explanation is structural, and phase-dependent. Writing $\theta'=\theta\cup u$, the choice of $g$ proportional to the likelihood gain of the added component,
\begin{equation}
\label{eq:gstar}
g^\star(u\,|\,\theta)\;\propto\;\frac{\mathcal{L}(\theta\cup u)}{\mathcal{L}(\theta)}\,p(u),
\end{equation}
renders the acceptance ratio \emph{independent of} $u$ at fixed prior and selection probabilities --- every proposed component is then accepted with the same probability, removing all $u$-variance of the acceptance; this is the sense in which we call it optimal. Through $\mathcal{L}$, $g^\star$ depends on the current residual $r=d-h(\theta)$ and therefore on the state $\theta$. The flow of Eq.~\eqref{eq:flow} is by construction state-independent: trained on the marginal distribution of wavelets already in the fit, it proposes where components already are, rather than where the fit is still deficient --- so during assembly it cannot compete with a residual-matched $g$, and no amount of training closes that gap (a perfectly trained unconditional flow still has the wrong conditioning structure). At equilibrium we argue the situation inverts: the residual is noise-dominated, and averaged over posterior states $g^\star$ approaches the posterior single-wavelet marginal --- the flow's training objective. We state the required assumption explicitly: the identification holds to the extent that components are nearly exchangeable and weakly dependent at the occupancies reached here, so that the state-averaged conditional for one more component tracks the population marginal; the state-resolved $g^\star$ retains state dependence (in particular, occupied regions remain blocked and the conditional for a supernumerary component concentrates at low amplitude), and we present the degeneracy as a structural argument supported by the controlled benchmark, not as a theorem. This is why flow births fire the $p(D)$-stationarity criterion in Table~\ref{tab:bench} while leaving the acceptance untouched, and why the same flow used as an in-model independence move (isolation row) buys nothing: within-model refinement is already served by tuned local moves.

We state the scope of this argument carefully. It is a statement about state-independent proposals, not about learned proposals in general, and it is not a theorem about the impossibility of learned trans-dimensional kernels. Eq.~\eqref{eq:gstar} identifies the object such a kernel would have to approximate, and a flow conditioned on a sufficient summary $\sigma(r)$ of the residual --- $g_\varphi(u\,|\,\sigma(r))$ --- would have the correct conditioning structure and remain valid under the reversible-jump~\cite{Green1995,Green2003} or involutive-MCMC~\cite{Neklyudov2020} constructions. The practical question, which we leave open, is whether such a conditional proposal can be trained online cheaply enough to outperform the closed-form residual matched filter that already serves as $g$.

\subsection{Software interface}
\label{sec:interface}
A new application requires two ingredients: a likelihood (vectorized over walkers) and a per-component prior box; everything else --- the reversible-jump machinery, the adaptive flow, its exact Green correction and its online retraining --- is supplied by the package. The image-counting experiment of Fig.~\ref{fig:pulses2d}, for example, reduces to
\begin{quote}
\begin{verbatim}
from eryn.ensemble import EnsembleSampler
from hyperwave import AdaptiveFlowProposal, FlowTrainingCallback, \
                      make_flow_rj_move

def log_like(params, groups):   # any likelihood over the D active
    ...                         # components of each walker

flow = AdaptiveFlowProposal(priors)             # learned birth proposal
sampler = EnsembleSampler(
    nwalkers, {"component": ndim}, log_like, priors,
    rj_moves=make_flow_rj_move({"component": flow},
                               nleaves_max={"component": 40}),
    update_fn=FlowTrainingCallback({"component": flow}),
    update_iterations=100, vectorize=True)
sampler.run_mcmc(start, nsteps)
\end{verbatim}
\end{quote}
with all defaults as in this paper (the flow architecture, retraining schedule, defensive prior mixture and stopping rule of this section). The wavelet, LISA, EEG and image applications differ only in \texttt{log\_like} and the prior box.

\subsection{Stopping rule}
\label{sec:converge}
Rather than a fixed step count we use a pre-specified stopping rule, fixed before any of the runs reported here and applied identically to every configuration. We stop when three conditions hold simultaneously on the cold chain: the sample length exceeds $A_\tau=50$ integrated autocorrelation times, $N > A_\tau\,\hat\tau$; the effective sample size exceeds a target, $\mathrm{ESS}=N N_{\rm walk}/\hat\tau > \mathrm{ESS}_{\min}=3000$; and the model-order distribution is stationary, measured by the total-variation distance between $p(D)$ estimated on the two halves of the retained chain, $\mathrm{TV}<0.02$. Reported run times are wall-clock to satisfy all three.

\paragraph{Diagnostics.} For a scalar chain $x_t$ with normalised autocorrelation function $\rho(k)$, the integrated autocorrelation time $\tau = 1 + 2\sum_{k\geq1}\rho(k)$ is the number of steps over which the chain decorrelates: $N$ retained steps of $N_{\rm walk}$ walkers carry $\mathrm{ESS}=N N_{\rm walk}/\tau$ independent-equivalent samples, which is what controls the Monte Carlo error of posterior estimates and, per unit wall time, the ESS/min of Table~\ref{tab:bench}. We estimate $\hat\tau$ with the FFT-based estimator with Sokal's automatic windowing~\cite{Sokal1997}, as popularised by \texttt{emcee}~\cite{ForemanMackey2013} and implemented in \textsc{Eryn}, averaged over walkers after discarding the first $30\%$ of the stored chain, on two scalars separately --- the cold-chain log-likelihood and the model order $D$ --- taking the more conservative $\hat\tau=\max(\hat\tau_{\log L},\hat\tau_{D})$. The estimate must additionally be stable between successive checks (relative change $<0.05$), and the full criterion must hold at two consecutive checks (performed every $200$ steps) before the run stops. Because walkers in an ensemble sampler are coupled by construction, the product $N N_{\rm walk}/\hat\tau$ treats inter-walker information as independent and should be read as an upper bound on the true effective sample size; it is computed identically for every configuration, so rankings and ratios between configurations are unaffected. Time-resolved autocorrelation times (Fig.~\ref{fig:phases}c) apply the same estimator within sliding windows of fixed length, so their values are not subject to the growing-window length dependence. Quantiles of the discrete model order $D$ shown in figures are computed after a continuity correction --- on the jittered variable $D+U(-\tfrac12,\tfrac12)$ --- so that credible bands are not pinned to integers.

We stress that no diagnostic can certify convergence of an MCMC chain; the rule above is a decision procedure, not a proof, and we use it only to compare configurations on an equal footing. $A_\tau=50$ is the standard heuristic for reliable $\hat\tau$ estimation in ensemble samplers~\cite{Goodman2010,ForemanMackey2013}.

\paragraph{The stationarity threshold carries a sample-size floor.} The two-half $\mathrm{TV}$ of $p(D)$ is a noisy statistic whose expectation under perfect stationarity is set by the effective sample size: across $54$ benchmark-family runs we measure a median $\mathrm{TV}$ that decays approximately as $1/\sqrt{N}$ and sits at $0.06$ at $N=24\,000$ steps --- so a threshold of $0.02$ is \emph{unreachable} at that length for any proposal, and only becomes resolvable at the $\sim\!50$--$70$k-step horizon at which the benchmark of Table~\ref{tab:bench} operates. A stopping threshold on a model-order stationarity statistic must therefore be checked against its own sample-size floor before it is used to compare samplers --- a threshold calibrated on a handful of runs can silently sit below the floor of the budget it is applied at, turning the binary ``converged'' outcome into a lottery over noise fluctuations. We recommend, and release with the package, a floor-calibrated variant of the rule (pass when the observed $\mathrm{TV}$ falls below a multiple of its stationary bootstrap expectation at the current effective sample size).

\subsection{Reproducibility of sampler benchmarks}
\label{sec:repro}
Two further reproducibility lessons from this study are worth recording, because either can silently invert the outcome of a sampler comparison. First, \emph{ensemble-sampler benchmarks must seed every random stream}: in the implementation used here the injection and initialization are seeded, but the move kernels draw from an unseeded stream, so repeated runs of an identical configuration scatter substantially (windowed $\hat\tau_D$ varying by $\pm30\%$ at matched settings); with few seeds this scatter can manufacture or bury a real effect. All comparisons in this paper therefore use ten seeds per arm with the full distribution reported. Second, \emph{the execution environment is part of the experiment}: a routine cluster maintenance (node kernel and GPU-driver update) between two of our campaigns slowed the mixing of \emph{every} configuration by a factor of $\sim3$ at bit-identical likelihood values, invalidating cross-era comparisons of absolute step counts while preserving the ordering between proposals. Wall-clock and step-count claims in this paper come from single post-maintenance campaigns on identical NVIDIA V100 nodes, and we report the environment (driver, kernel, package versions) alongside the code.

\section*{Acknowledgments}
A.S.\ and M.W.C.\ acknowledge support from the National Science Foundation with grant numbers PHY-2308862 and PHY-2117997. N.~K.\ acknowledges funding from the European Union's Horizon 2020 research and innovation program under the Marie Sk\l{}odowska--Curie grant agreement No.~101065596. N.~K.\ was supported by the
Hellenic Foundation for Research and Innovation (H.F.R.I.) under the 4th Call for HFRI Research Projects to support Post-doctoral Researchers (Project Number 28418). N.~S.\ and N.~K.\ acknowledge support from the Gr-PRODEX 2019 funding program (PEA 4000132310). N.~S.\ acknowledges funding from the H.F.R.I.\ Project
No.~26254. In addition, this publication is part of a project that has received funding from the European Union's Horizon Europe Research and Innovation Programme under Grant Agreement No.~101131928. M.K.\ acknowledges funding from NSF Award Number 2311559, and from the U.S.\ Department of Energy, Office of Science, Office of Advanced Scientific Computing Research under Contract No.~DE-AC02-05CH11231 at Lawrence Berkeley National Laboratory to enable research for Data-intensive Machine Learning and Analysis. We are grateful for the
computational resources provided by the Minnesota Supercomputing Institute (MSI) at the University of Minnesota.

\section*{Author contributions}
A.S.\ conceived and led the project: A.S.\ developed the phase-structure argument and the learned birth-proposal machinery, implemented the \hyperwave{} package, performed all experiments and analyses, and wrote the manuscript. N.~K.\ is a co-developer of \textsc{Eryn} and \hyperwave{}, and contributed comments and improvements to the manuscript. M.K.\ is a co-developer of \hyperwave{} and \textsc{pocoMC}, and contributed comments and improvements to the manuscript. M.L.K.\ is a co-developer of \textsc{Eryn}. D.K.\ independently reviewed the \hyperwave{} package and reproduced the reported runs. All authors reviewed and approved the final manuscript.

\section*{Competing interests}
The authors declare no competing interests.

\section*{Data availability}
All data are public. GW150914 strain is from the Gravitational Wave Open Science Center~\cite{GWOSC}. The LISA massive-black-hole-binary signal is injected from the \textsc{Sangria} (LDC2a) parameters of the LISA Data Challenge~\cite{LDC}. The electroencephalogram is record \texttt{chb01\_03} (channel FP2--F4) of the CHB-MIT Scalp EEG Database on PhysioNet.

\section*{Code availability}
The open-source \hyperwave{} package, and scripts reproducing every figure and table, are available at \url{https://github.com/asasli/HyperWave}. \hyperwave{}
builds on the \textsc{Eryn} parallel-tempered RJMCMC sampler~\cite{Karnesis2023}; the contributions introduced here --- the GPU-resident batched wavelet likelihood, the adaptive normalizing-flow birth proposal with its exact Green correction, the shape-informed
guided birth moves, the automatic real-strain data conditioning (Methods~\ref{sec:conditioning}), and the stopping rule --- are provided as part of \hyperwave{}.

\bibliographystyle{unsrt}
\bibliography{refs}

\end{document}